\documentstyle[11pt,newpasp,twoside,epsf]{article}
\DeclareRobustCommand{\ion}[2]{%
\relax\ifmmode
\ifx\testbx\f@series
{\mathbf{#1\,\mathsc{#2}}}\else
{\mathrm{#1\,\mathsc{#2}}}\fi
\else\textup{#1\,{\mdseries\textsc{#2}}}%
\fi}

\def\edcomment#1{\iffalse\marginpar{\raggedright\sl#1\/}\else\relax\fi}
\reversemarginpar

\begin{document}
\title{Helium-like ions as powerful X-ray plasma diagnostics}
 \author{D. Porquet}
\affil{CEA/DSM/DAPNIA Service d'Astrophysique F-91191 Gif-sur-Yvette, France}
\author{R. Mewe, A.J.J Raassen, J.S. Kaastra}
\affil{SRON Laboratory for Space Research Sorbonnelaan 2, 3584 CA Utrecht, The Netherlands}
\author{J. Dubau}
\affil{Observatoire de Paris, UPR 176 CNRS, DARC F-92195 Meudon-Cedex, France}

\begin{abstract}
We  revisited the calculations of the ratios of the Helium-like ion ``triplet'' ({\it resonance}, {\it intercombination}, and {\it forbidden} lines)  for Z=6 to 14 (\ion{C}{v}, \ion{N}{vi}, \ion{O}{vii}, \ion{Ne}{ix}, \ion{Mg}{xi}, \ion{Si}{xiii}) in order to provide  temperature, density and ionization  diagnostics for the new high-resolution spectroscopic data of {\sl Chandra} and {\sl XMM-Newton}. Comparing to earlier computations, collisional rates are updated and the best experimental values for radiative transition probabilities are used. The influence of an external  radiation field (photo-excitation) and the contribution from unresolved dielectronic satellite lines to the line ratios are discussed. Collision-dominated plasmas (e.g. stellar coronae), photo-ionized plasmas (e.g. AGNs) or transient plasmas (e.g. SNRs) are considered.
\end{abstract}

\section{Introduction}
The new generation of X-ray satellites (Chandra, XMM-Newton) enable us to obtain unprecedented high spectral resolution and high S/N spectra. The wavelength ranges of the {\sl LETGS}\footnote{LETGS: Low Energy Transmission Grating Spectrometer on Chandra.} (6--170 \AA) and the {\sl RGS}\footnote{RGS: Reflection Grating Spectrometer on XMM-Newton.} (6--35 \AA) contain the Helium-like line ``triplets'' from \ion{C}{v} (or \ion{N}{vi} for the RGS) to \ion{Si}{xiii}. The triplet consists of three close lines: the {\it resonance} line, the {\it intercombination} line and the {\it forbidden} line.  The Helium-like triplets provide electron density ($n_e \sim$~10$^8$--10$^{13}$~cm$^{-3}$) as well as electron temperature ($T \sim 1$--10~MK). The line ratios of these He-like triplets enable also the determination of the ionization processes (photo-ionization and/or collisional ionization) which prevail in the plasma.\\
\indent The ratios of these lines are already widely used for collisional solar plasma diagnostics (e.g., Gabriel and Jordan 1969, Doyle 1980, Keenan et al. 1987, McKenzie \& Landecker 1982). \\
Recently, also theoretical calculations for photo-ionized plasmas or ``hybrid'' plasmas (photo-ionization plus collisional ionization) have been  given by Porquet \& Dubau (2000). Their calculations have been already applied to spectra of Seyfert galaxies (NGC\,5548, Kaastra et al. 2000; Mkn\,3, Sako et al. 2000; NGC\,4151, Ogle et al. 2000).\\
 We  present here calculations of these ratios, from \ion{C}{v} to \ion{Si}{xiii}, which could be applied directly for the first time to {\sl Chandra} and {\sl XMM-Newton} observations of the extra-solar collisional plasmas such as stellar coronae. These calculations have been done to apply an improved model to the density analysis of the {\sl LETGS} and {\sl RGS} spectra of various late-type stars such as Capella, Procyon, and $\alpha$~Centauri (e.g., Audard et al. 2001, Ness et al. 2001, Mewe et al. 2001a) and also to O stars such as Zeta Puppis (Kahn et al. 2001). Our model is to be considered as an improvement of various previous calculations for solar plasmas such as done by e.g., Gabriel \& Jordan (1969), Mewe \& Schrijver (1978), Pradhan \& Shull (1981), Mewe et al. (1985), and Pradhan (1985). The calculations are based on recent work by Porquet \& Dubau (2000) and by Mewe et al. (2001b). \\
\indent In the next two sections, we introduce the plasma diagnostics and the atomic processes taken into account in the calculations. In sect. 4 we discuss applications to different types of plasmas (photo-ionized, collision-dominated, and transient plasmas).
\section{Plasma diagnostics using Helium-like ions}

\indent In the X-ray range, the three most intense lines of Helium-like ions (``triplet'') are: the {\bf resonance} line ($r$, also called $w$: 1s$^{2}$\,$^{1}S_{\mathrm{0}}$ -- 1s2p\,$^{1}P_{\mathrm{1}}$, ), the {\bf intercombination} lines ($i$, also called $x+y$: 1s$^{2}$\,$^{1}S_{0}$ -- 1s2p\,$^{3}P_{2,1}$) and the {\bf forbidden} line ($f$, also called {\bf z}: 1s$^{2}$\,$^{1}{S}_{0}$ -- 1s2s\,$^{3}{S}_{1}$). They correspond to transitions between the $n$=2 shell and the $n$=1 ground state shell (see Figure~1). The energy of each line from \ion{C}{v} (Z=6) to \ion{Si}{xiii} (Z=14) are reported in Table~1.\\

\begin{figure}
\plotfiddle{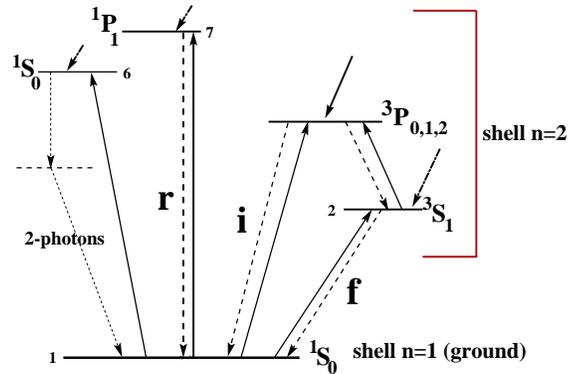}{5.5cm}{}{40}{40}{-100}{-100}
\caption{Simplified Gotrian diagram of He-like ions.}
\label{Gotrian}
\end{figure}

As shown by Gabriel \& Jordan (1969), these ratios are sensitive to electron density and to the electron temperature:
\begin{equation}
R~(n_e)~=~\frac{f}{i}~~~~~ \left({\mathrm{also~~}} \frac{z}{x+y}\right) \\
\label{eq:R}
\end{equation}
\begin{equation}
G~(T_e)=\frac{f+i}{r}~~~~~  \left({\mathrm{also}}~~ \frac{z+(x+y)}{w}\right)  \\
\label{eq:G}
\end{equation}

At low density, where the n=2 states are populated by electron excitation and then decay radiatively, the relative intensities of the three lines are  independent of the density. Above the {\it critical density}\footnote{The {\it critical density} (n$_{crit}$) is defined as n$_{crit}$C$\sim$A: C being the rate coefficient for collisional excitation 2$^{3}$S $\to$ 2$^3$P, A being the radiative transition probability of 2$^{3}$S $\to$ 1$^1$S.}, the level 2$^{3}$S (upper level of the {\it forbidden} line) becomes to be depleted by collisions to the  2$^{3}$P levels (upper levels of the intercombination line). Then, at increasing density, the intensity of the {\it forbidden} line strongly decreases while the intensity of the {\it intercombination} line increases, implying a lowering of the ratio $R$ over approximatively two decades of density (see figure~8 in Porquet \& Dubau 2000). However caution should be taken for \ion{C}{v} and \ion{N}{vi} since in case of an intense UV radiation field, the photo-excitation between the $^{3}$S term and the $^{3}$P term  is not negligible and has the same effect on the {\it forbidden} line and on the {\it intercombination} line as the collisional coupling, and thus could mimic a high-density plasma.\\
\indent The ratio $G$ is sensitive to the electron temperature since the collisional excitation rates have not the same dependence with the temperature for the {\it resonance} line and the two other lines.\\
\indent Finally, the relative intensity of the three lines could be used to determine the ionization processes which prevail in the plasma, i.e. photo-ionization and/or collisional ionization (see Porquet \& Dubau 2000, and the references therein). In the case of a plasma dominated by  photo-ionization and recombination, the {\it forbidden} line (or the {\it intercombination} line at high density) is much stronger than the {\it resonance line} (hence $G \gg 1$), whereas in a plasma dominated by collisional ionization and excitation the {\it resonance} line is comparable or stronger than the two other ones ($G \sim 1$) (see figure~11 in Porquet \& Dubau 2000).

\begin{table}
\label{lambda}
\caption{Energy (in eV) of the three main X-ray lines of \ion{C}{v}, \ion{N}{vi}, \ion{O}{vii}, \ion{Ne}{ix}, \ion{Mg}{xi} and \ion{Si}{xiii}, as well as the corresponding wavelength in \AA, in parentheses.}
\vspace{-0.4cm}
\begin{center}
\begin{tabular}{ccccccc}
\hline
\hline
            \\  
\small{line}                 & \small{\ion{C}{v}}&\small{\ion{N}{vi}}&\small{\ion{O}{vii}}  &\small{\ion{Ne}{ix}} &\small{\ion{Mg}{xi}}  &\small{\ion{Si}{xiii}}\\
            \\  
\hline
\small{resonance (r)}        & \small{307.88}    & \small{430.65}    & \small{574.00}       &  \small{921.82}      & \small{1357.07}     & \small{1864.44 }   \\  
                     &\small{(40.27\AA)}    &\small{(28.79\AA)}    &\small{(21.60\AA)}       & \small{(13.45\AA)}      &\small{(9.17\AA)}       & \small{(6.65\AA)} \\
\small{intercombination (i)} & \small{304.41}    & \small{426.36}    & \small{568.74}       & \small{915.02}       &\small{1343.28}      & \small{1853.29}    \\  
                     &\small{(40.73\AA)}    &\small{(29.08\AA)}    &\small{(21.80\AA)}       &\small{(13.55\AA)}       &\small{(9.23\AA)}       &\small{(6.69\AA)} \\
\small{forbidden (f)}        & \small{298.97}    & \small{419.86}    & \small{561.02}       & \small{905.00}       & \small{1331.74}     & \small{1839.54}    \\  
                     & \small{(41.47\AA)}   &\small{(29.53\AA)}    &\small{(22.10\AA)}       &\small{(13.70\AA)}       &\small{(9.31\AA)}      &\small{(6.74\AA)} \\
\hline
\hline
\end{tabular}
\end{center}
\vspace{-0.4cm}
\end{table}

\section{Atomic processes}

\indent The rates for the collisional excitations and for the radiative and dielectronic recombinations are from the recent data from Porquet \& Dubau (2000), see also references therein.\\
\indent The A$_{ki}$ for the {\it forbidden} line (2$\to$1), and for the strongest component of the {\it intercombination} line (4$\to$1) are updated by their experimental values when available (see Mewe et al. 2001b).\\
\indent A strong radiation field could mimic a high-density plasma analogously to collisional coupling above the critical density.  This effect can be important for \ion{C}{v} and also for \ion{N}{vi} (cf. also Ness et al. 2001), but for higher-Z ions this effect is usually negligible. For the formulae of the photo-excitation rates, see Mewe et al. (2001b).\\
\indent In some cases where the plasma column density is high, optical depth effects on the resonance line should be taken into account in the calculation of the ratio $G$. Determination of the ratio of the possibly optically thick resonance line to the optically thin forbidden line was already suggested by Acton and Brown (1978) as a diagnostic to investigate resonance scattering effects in the solar corona. Resonant scattering could be important in plasmas such as the ``Warm Absorber'' in AGNs where the column density $N_{H}$ could be as high as 10$^{21}$--10$^{23}$\,cm$^{-2}$, and also in some cases in stellar coronae.\\
\indent Moreover, the contributions of unresolved dielectronic satellite lines to the  lines {\it r} and {\it f} for \ion{Ne}{ix}, \ion{Mg}{xi} and \ion{Si}{xiii} are taken into account.

\section{Applications to astrophysical plasmas}

        \subsection{Photo-ionized and ``hybrid'' plasmas: Warm Absorber in AGNs}

\indent The Warm Absorber (WA) is a totally or a partially photo-ionized medium (with or without an additional ionization process) located in the central part of Active Galactic Nuclei. It may be a warm or hot component of the ``Broad Line Region'' and/or the ``Narrow Line Region''. The ionization processes that occur in the WA, are still not very well known. Indeed, even though the WA is commonly thought to be a photo-ionized gas, an additional ionization process cannot be ruled out (Porquet \& Dumont 1998, Porquet et al. 1999, Nicastro et al. 1999). Thus, in Porquet \& Dubau (2000), both types of plasmas have been taken into account: a ``purely photo-ionized plasma'' and a ``hybrid plasma'' which is a photo-ionized plasma with additional collisional ionization processes. Porquet \& Dubau (2000) have calculated the ratios $R$ and $G$ for both types of plasmas (see also the atomic data therein).\\
\indent Recently, {\sl Chandra} provides much better spectral observations of the WA in AGNs, which show clearly the presence of He-like emission lines. For NGC\,5548, Kaastra et al. (2000) have deduced from the line ratio of the triplet of \ion{O}{vii} that the plasma is photo-ionized and they have also obtained an upper limit to the density ($<$7.10$^{10}$\,cm$^{-3}$),  also in the Seyfert 2 Mkn 3  \ion{O}{vii} is produced by a plasma where  photo-ionization prevails (Sako et al. 2000). For the Seyfert NGC\,4151 (type 1.5), Ogle et al. (2000) have inferred from the He-like ion triplets that the WA is a ``hybrid'' plasma where both photo-ionization and collisional ionization occur.\\
\indent Improvements of these calculations for photo-ionization plasmas are in preparation, taking into account experimental values for the transition probabilities of the $forbidden$ and the $intercombination$ lines, and the effect of an external radiation field (photo-excitation) for \ion{C}{v} and \ion{N}{vi}, as well as the contributions  of unresolved dielectronic satellite lines to the $resonance$ and the $intercombination$ line, which  for a low-temperature plasma  could not be negligible.

\subsection{Collision-dominated plasmas: stellar coronae}
\indent Recently, new line ratio calculations have been made for collision-dominated plasmas in Mewe et al. (2001b). Comparing to earlier computations, collisional rates are updated and the best experimental values for radiative transition probabilities are used. The influence of an external stellar radiation field (photo-excitation) and the contribution from unresolved dielectronic satellite lines are taken into account.\\
These ratio calculations have been recently used by Audard et al. (2001), Ness et al. (2001), and Mewe et al. (2001a) for applications to spectral data obtained with the {\sl Chandra-LETGS} and {\sl RGS} for Procyon and Capella (late-type stars).\\
\indent Several works have been already dedicated to the study of Capella (Audard et al. 2001, Canizares et al. 2000, Brinkman et al. 2000, Mewe et al. 2001a, Ness et al. 2001) and  Procyon (Ness et al. 2001). Density and temperature diagnostics based on the He-like ions have been used. The inferred values show that the corona of Capella is bimodal with a low-density cool plasma and hotter plasma with high densities. Then on the basis of the derived temperatures and densities, typical loop length scales of the star can be estimated using the known loop scaling law derived by Rosner et al. (1978).\\
Recently, Kahn et al. (2001) have found with {\it XMM-Newton} that for $\zeta$ Puppis, the $forbidden$ to $intercombination$ line ratios within the Helium-like triplets are abnormally low for \ion{N}{vi}, \ion{O}{vii}, and \ion{Ne}{ix}. While this is sometimes indicative of high electron density, they have shown that in the case of $\zeta$ Puppis, it is instead caused by the intense radiation field of this star. This constraints the location of the X-ray emitting shocks relative to the star, since the emitted regions should be close enough in order that the UV radiation is not too much diluted. 
\subsection{Transient plasmas: SNRs}
\indent Recently, line-rich, high-resolution X-ray spectra were obtained with the {\sl XMM-RGS} on two bright supernova remnants (SNRs) located in the Small Magellanic Cloud, i.e. 1E 0102-72.3 (Rasmussen et al. 2001) and N132D (Behar et al. 2001). Because these sources are expected to be in a highly transient ionizing state it is important to take into account the effects of non-equilibrium ionization (NEI) in the interpretation of the line ratios.
In the currently publicly available Utrecht spectral code SPEX 1.10 (Kaastra et al. 1996) several NEI SNR models are incorporated with calculations of the He-like triplet lines from Mewe et al. (1985) which can be used as a first approximation to investigate the NEI effects from an analysis of the well-resolved lines of e.g., the \ion{O}{vii} triplets. In a later version of SPEX our new calculations will be implemented.


\section{Conclusion and perspectives}
For the first time, thanks to the new generation of X-ray satellites, {\sl Chandra} and {\sl XMM-Newton} the diagnostics based on the line ratios of He-like ion could be used for powerful extra-solar plasma diagnostics (Warm Absorber in AGNs, stellar coronae, ...). These diagnostics are one of the keys for a better understanding of the solar-stellar connection: heating of the coronae, magnetic activity, etc. The applications of these diagnostics also allow to discriminate between a photo-ionized and a ``hybrid'' plasma in different types of AGNs. This is crucial in order to better understand the physical parameters of the accreting matter around a black hole in AGNs, and hence the evolution of active galaxies.


\begin{references}
\small{
Acton L.W. \& Brown W.A. 1978, ApJ, 225, 1065\\
Audard M., Behar E., G{\"u}del M. et al. 2001, A\&A, 365 {\it special XMM-Newton issue}, in press\\
Behar E, Rasmussen A.P., Gareth Griffiths R. et al. 2001, A\&A, 365 {\it special XMM-Newton issue}, in press\\
Brinkman A.C., Gunsing C.J.T., Kaastra J.S. et al. 2000, \apjl, 530, L111 \\
Canizares C.R., Huenemoerder D.P., Davis D.S. et al. 2000, \apjl, 539, L41 \\
Doyle J. G. 1980, \aap, 87, 183 \\
Gabriel A. H., Jordan C. 1969, \mnras, 145, 241 \\
Kaastra J.S., Mewe R., Liedahl D.A. et al. 2000, \aap, 354, L83 \\
Kaastra J.S., Mewe R., Nieuwenhuijzen H., 1996, in UV and X-ray Spectroscopy of Astrophysical and Laboratory Plasmas, K. Yamashita and T. Watanabe (eds.), 
Tokyo, Universal Academy Press, p. 411 (SPEX)\\
Kahn S.M., Leutenegger M.A., Cottam J. et al. 2001, A\&A, 365 {\it special XMM-Newton issue}, in press\\
Keenan F.P., McCann S.M., Kingston A.E., McKenzie D.L. 1987, \apj, 318, 926\\
McKenzie D.L., Landecker P.B. 1982, \apj, 259, 372\\ 
Mewe R. 1972, \solphys, 22, 114\\
Mewe R., Gronenschild E.H.B.M., van den Oord G.H.J. 1985, A\&AS, 62, 197\\
Mewe R. \& Schrijver J. 1978, \aap, 65, 99 \\
Mewe R., Raassen A.J.J., Drake J. et al. 2001a, A\&A, submitted\\
Mewe R., Porquet D., Raassen A.J.J., Kaastra J., Dubau, J. 2001b, A\&A, in prep.\\
Ness J.-U., Mewe R., Schmitt J.H.M.M. et al. 2001, A\&A, submitted\\
Ogle P.M., Marshall H.L., Lee J.C., Canizares C. R. 2000, ApJL, in press\\ 
Porquet D. \& Dumont A.-.M 1999, ASP Conf.\ Ser.\ 175, 359 \\
Porquet D., Dumont A.-M., Collin S., Mouchet M.\ 1999, A\&A, 341, 58\\ 
Porquet D. \& Dubau J. 2000, \aaps, 143, 495\\ 
Pradhan A.K. \& Shull 1981, \apj, 249, 821\\ 
Pradhan A.K. 1985, \apj, 288, 824 \\
Rasmussen A.P., Behar E., Kahn S.M. et al. 2001, A\&A, 365 {\it special XMM-Newton issue}, in press\\
Rosner R., Tucker W.H., Vaiana G.S. 1978, ApJ, 220, 643 \\
Sako M., Kahn S.M., Paerels F. \& Liedahl D.A. 2000, ApJ, 543, L115 }
\end{references}
\end{document}